\documentclass[a4paper,11pt]{article}
\usepackage{pos}
\pdfoutput=1

\title{
The $Z_{cs}(3985)$ structure,
if a triangle/kinematic singularity, would disappear when heated in Heavy Ion Collisons.}
 \ShortTitle{$Z_{cs}(3985)$ disappearance in heavy ion collisions if a triangle singularity}

\author*[a]{Felipe J. Llanes-Estrada}
\author[b]{Luciano M. Abreu}

\affiliation[a]{Univ. Complutense de Madrid, Dept. F\'{\i}sica Te\'orica, Plaza de las Ciencias 1, 28040 Madrid, Spain}

\affiliation[b]{Instituto de F\'isica, Universidade Federal da Bahia, Salvador, Bahia, 40170-115, Brazil}

\emailAdd{fllanes@fis.ucm.es}
\emailAdd{luciano.abreu@ufba.br}

\abstract{Triangle and other kinematic singularities are very sensitive to the precise masses and widths of the intervening particles. Therefore, the effect that the heavy-ion collision medium can have on those masses and widths, as captured by finite-temperature field theory and reported in the literature, 
may erase the singularity from the spectrum if the effect is large enough and the loop completes before the hot gas freezes out. A very timely example is provided by the $Z_{cs}(3985)$ structure recently reported by BES-III in a $(D_s^-D^{*0}+D_s^{*-}D^0)$ spectrum recoiling against a $K^+$. If a new hadron, this would be a clear  exotic $c\bar{c}s\bar{u}$ tetraquark candidate, the first of a charmonium-like family with strangeness: to accept this, alternative explanations first need to be tested and discarded. As shown in figure 1, the mass spectrum recoiling against the kaon is near the $m_{D_s^{*-}}+m_{D^0}$ threshold, whereas
the production cross-section seems to peak around that for $\sqrt{s(e^-e^+)}=m_{D_{s2}^{*+}} +m_{D_s^{*-}} $. Therefore, this structure may well be caused by a triangle with the three charmed mesons 
$D_{s2}^{*+}  /  D_s^{*-} / D^0$ running in the loop, with the amplitude enhanced at the two thresholds. If so, the structure would be erased from the spectrum in Heavy Ion Collisions, whereas a hadron would continue to exist although with a mass decreased by a few percent as computed by other authors.}

\FullConference{%
  *** The European Physical Society Conference on High Energy Physics (EPS-HEP2021), ***\\
  *** 26-30 July 2021 ***\\
  *** Online conference, jointly organized by Universität Hamburg and the research center DESY *** 
}

\begin{document}

\maketitle
%%%%%%%%%%%%%%%%%%%%%%%%%%%%%%%%%%%%%%%%%%%%%%%%%%%%%%%%%%%%%%%%%%%%%%%%%%%%%%%%%%%%%%%%%
The Heavy-Ion Collision program can make contributions to hadron spectroscopy impacting the classification of the many newly found hadron structures~\cite{Hu:2021gdg}. 
We have recently proposed that triangle singularities (and by extension, other types of Landau singularities that need kinematic accidents), increasingly common in hadron spectroscopy due to the sophisticated multiparticle analysis now available, can often be separated from ordinary or exotic hadrons because they are erased by the effect of a medium. In addition to the examples initially illustrated~\cite{Abreu:2020jsl}, we have shown~\cite{Abreu:2021xpz} how the method can be used to ascertain whether the $Z_c(3900)\to J/\psi\pi $ structure in $Y(4260)\to J/\psi\pi\pi$ decays, as proposed earlier~\cite{Wang:2013cya}, proceeds by a triangle singularity.

\begin{figure}[h] \centering
\includegraphics[width=0.47\columnwidth]{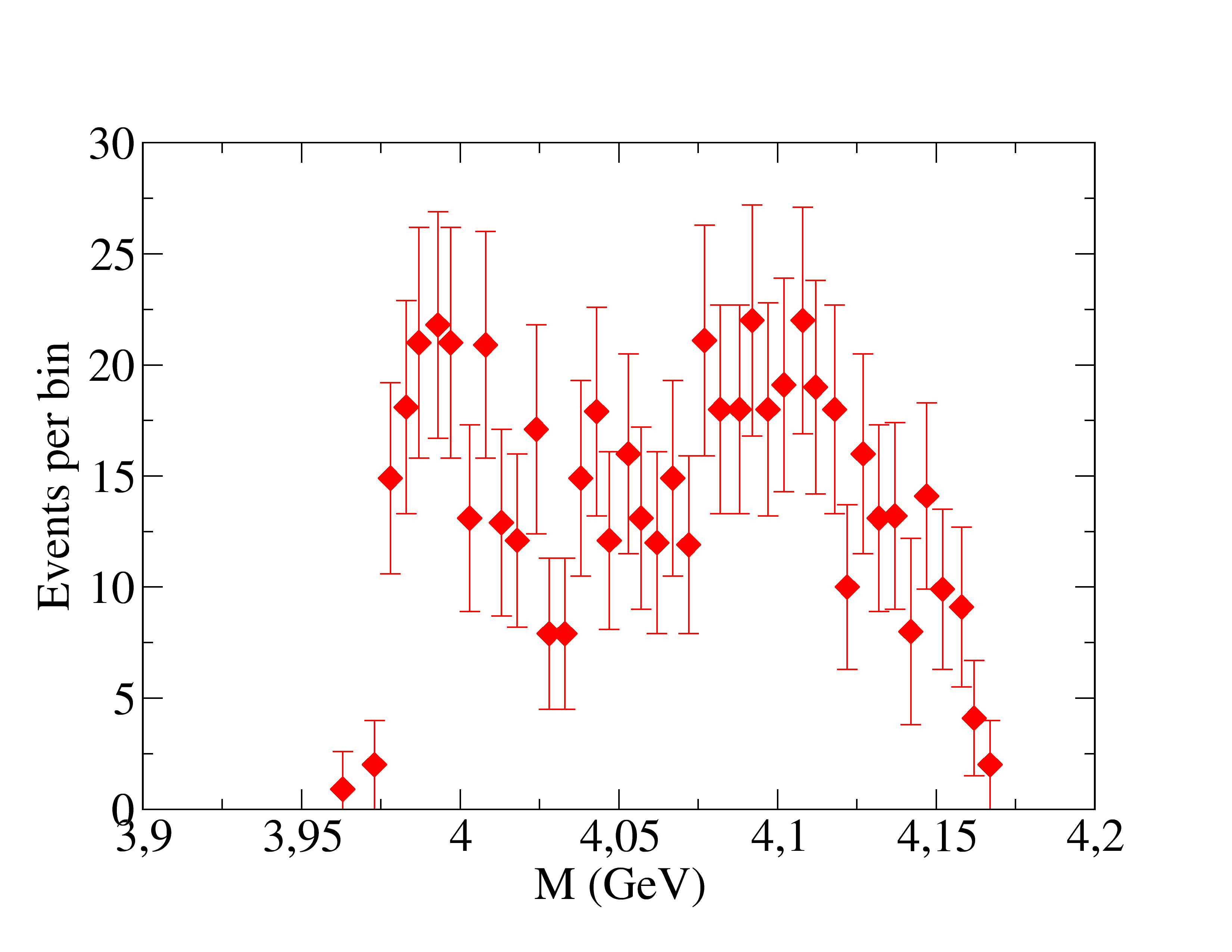}
\includegraphics[width=0.47\columnwidth]{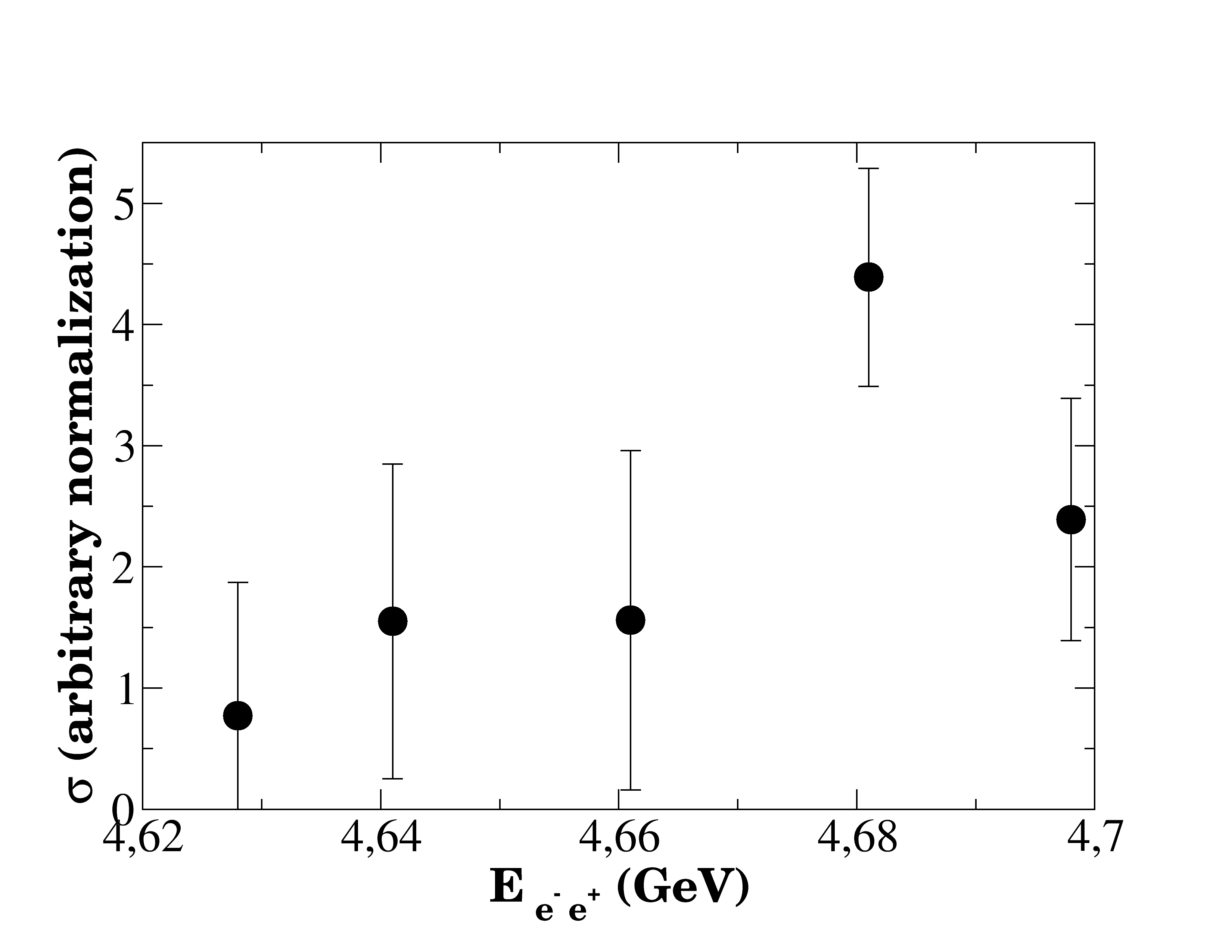}\\
\includegraphics[width=0.5\columnwidth]{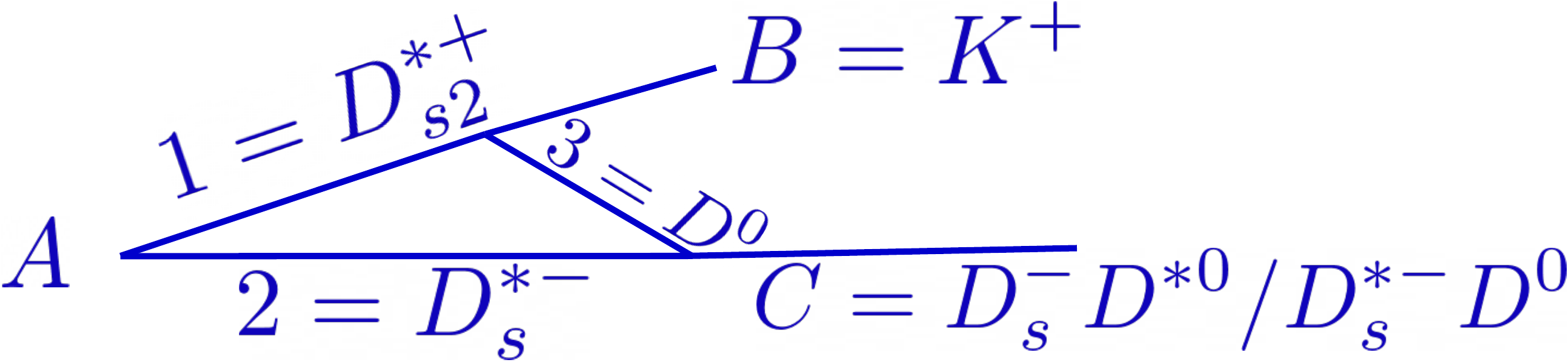}
\caption{Top left: The $(D_s^-D^{*0}+D_s^{*-}D^0)$ mass spectrum suggesting the existence of $Z_{cs}(3985)$. Top right: its production cross-section seems peaked at 4.68 GeV. 
Bottom: these two peaks are suggestive of the two thresholds (2+3) and (1+2), respectively, in the depicted triangle diagram.
Data from~\cite{BESIII:2020qkh}, newly rendered with symmetric error bars.   
\label{fig:data}}
\end{figure}

Here we turn to the $Z_{cs}(3985)$ structure reported by BES-III~\cite{BESIII:2020qkh} and shown in figure~\ref{fig:data}.
The community of hadron spectroscopy has been busy trying to ascertain whether it accepts a multiquark interpretation~\cite{Garcilazo:2021nyz}, whether one can expect a molecular composition~\cite{Sun:2020hjw}, or whether 
threshold and/or triangle production effects~\cite{Ikeno:2020mra} are at play and no new hadron is expected. 
To bring future heavy-ion collision data to bear, we point out in this contribution that the threshold/triangle kinematic coincidences are destroyed by the finite-temperature medium, whereas a real hadron is expected to continue existing in the medium though its mass is expected to drop perhaps 10\% at the finite temperature just below the phase transition between the quark-gluon plasma and the hadron gas~\cite{Sungu:2020zvk}, and similarly at finite density~\cite{Azizi:2020zyq}.

\begin{figure}
    \centering
    \includegraphics[width=0.49\columnwidth]{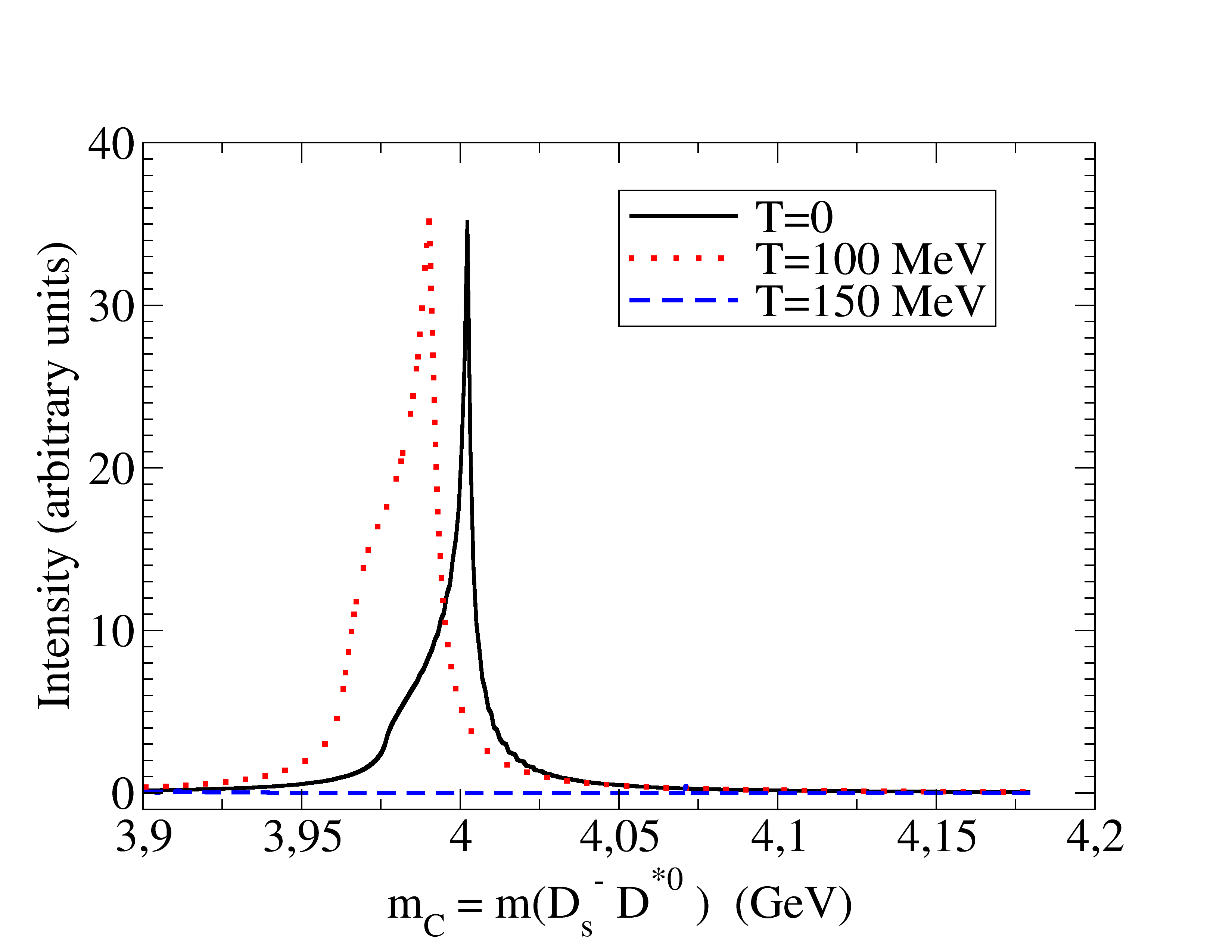}
    \includegraphics[width=0.49\columnwidth]{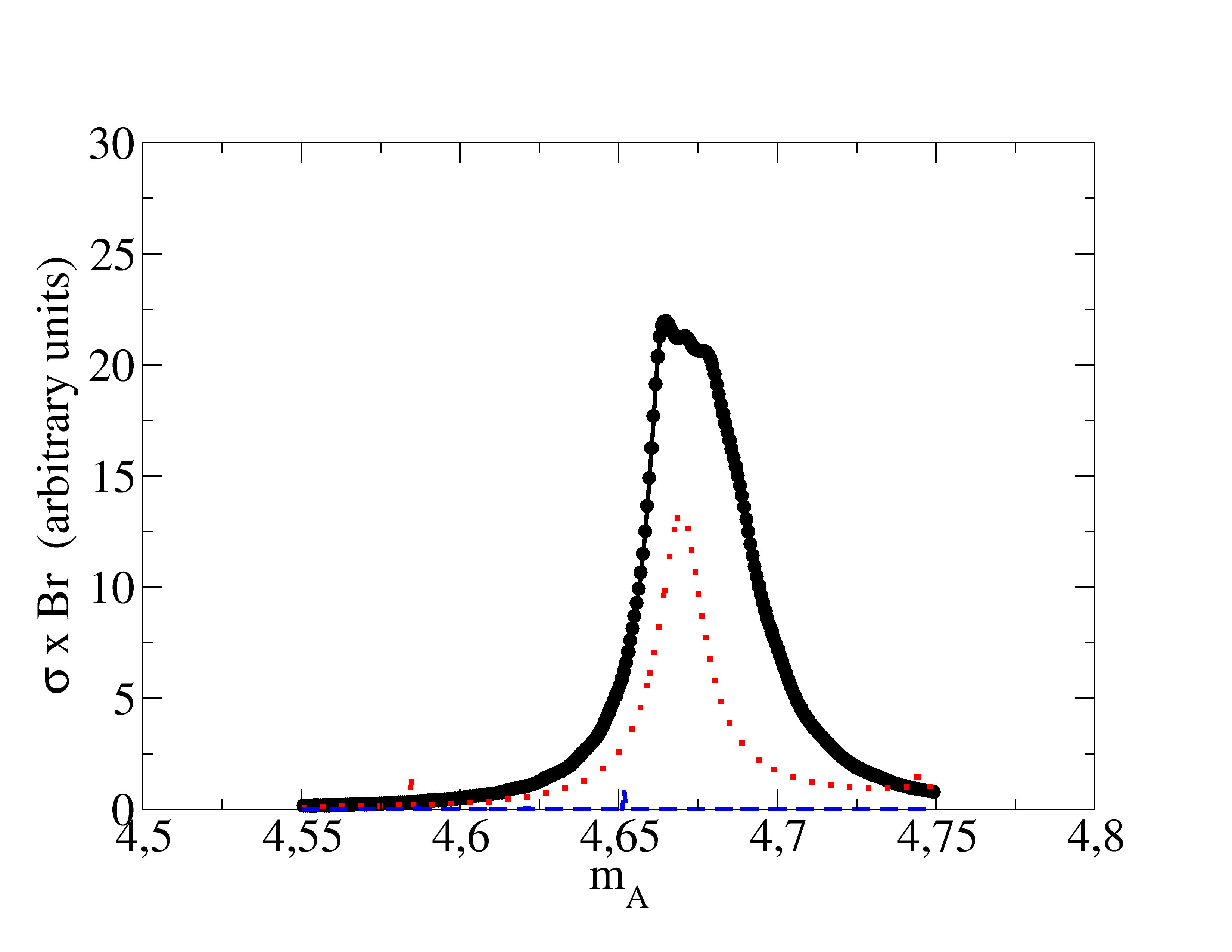}
    \caption{The left and right plots correspond to those in figure~\ref{fig:data}, respectively, but show the effect of the temperature with our understanding of the behavior of the masses and widths of the particles involved in the loop, as given in table~\ref{tab:params}. At 150 MeV the singularity is erased from the spectrum. }
    \label{fig:triangledisappears}
\end{figure}

To show that the triangle feature disappears at sufficient temperature we evaluate 
the triangle diagram in thermal field theory for an infinite, stationary hadron gas,
\begin{eqnarray}    
\label{loop-finiteT} 
I_{\triangleleft}  \simeq    \frac{1}{2} \int \frac{d^3 q}{(2 \pi)^3} \frac{1}{8 E_1 E_2 E_3} \frac{1}{\left(E_A - \tilde E_1 -\tilde E_2 \right)} 
  \frac{1}{\left( E_C - \tilde E_2 - \tilde E_3 \right)}   \frac{1}{\left( E_B - \tilde E_1 + \tilde E_3 \right)}  \times \nonumber \\   \nonumber
  \left\{ \left[ 1 + 2n_\beta(\tilde E_2) \right] \left(\! \tilde E_1 \!   -\! E_B  \! -\! \tilde E_3\! \right) 
+
\left[ 1+2n_\beta(E_A- \tilde E_1 ) \right] \left(\! E_C \! -\!  \tilde E_2\! -\!  \tilde E_3\! \right)  
+  \left[ 1+2n_\beta (\tilde E_3-E_C) \right] 
\left(\!E_A\! -\! \tilde E_1\! -\! \tilde E_2\! \right) \right\}\ .
\end{eqnarray}
The temperature enters through the Bose-Einstein factors $n_\beta$ but also through the in-medio mass and width of the three particles in the triangle
 (the width is contained in $\tilde E_\alpha = E_\alpha-i\frac{\Gamma_\alpha}{2}$). 
We have employed an estimate of the various $m_i$, $\Gamma_i$ based on available information in the literature~\cite{Veliev:2009tb,Montana:2020lfi}, shown in table~\ref{tab:params}. The result of this example computation is given in figure~\ref{fig:triangledisappears}.

\begin{table}[h]\centering
\caption{Parameters for the triangle computation at zero and finite temperature. 
At 100 MeV we decrease the $D^*_s$ masses by a plausible 5 MeV in view of~\cite{Montana:2020lfi}, while at 150 MeV the decrease is clearer. 
All units in MeV. (For the widths that are too narrow for the strong force, we adopt a $\Gamma=\epsilon=0.1$ MeV regulator).\label{tab:params}}
\begin{tabular}{|c|cccccc|} \hline
$T$ & $m_1=m_{D_{s2}^{*+}}$ & $m_2=m_{D_{s}^{*-}}$ & $m_3=m_{D^0}$ & $\Gamma_1$ & $\Gamma_2$ & $\Gamma_3$ \\ 
\hline
0   & $2568\pm 1$ & $2112.2\pm 0.4$ &  $1864.84\pm 0.05$ & $6.9\pm 0.7$ & $<2$ & $\approx$ stable \\
100 & 2563 & 2107 & 1856 & 12 & 2 & 0.8 \\ 
150 & 2500 & 2087 & 1776 & 15 & 10 & 2 \\
       \hline
\end{tabular}
\end{table}

The $T=0$ solid black line for both the invariant mass of the presumed particle at 3985 MeV (left) and for the $m_A$ variable relevant for the production cross-section  show distinctive peaks. We thus confirm that a singularity driven by two thresholds, one for $m_C$ and one for $m_A$, may be a satisfactory explanation of the excess in the data.

Proceeding to finite temperature, we see that at 100 MeV, the spectrum (left plot) is modestly shifted to smaller invariant mass and quite broadened, but the resonance is still clearly visible. 
The more significant changes of the masses and widths that occur at 150 MeV temperature entail that the structure is now clearly erased. 

The same phenomenon is seen in $|I_{\triangleleft}(m_A^2)|^2$, that controls the production cross-section, on the right plot: there is a substantial drop at 100 MeV, and complete disappearance at 150 MeV. We expect this to be even faster in a real heavy-ion collision given that we have not taken into account the folding with the energy spectrum in the initial state: whereas the $e^-e^+$ cm energy is precisely known, in a heavy ion collision the initial state contains colliding particles of all momenta. This will produce large nonpeaking backgrounds that will erase the structure even faster.

The question remains whether the triangle can be completed within the
time span of a heavy ion collision, before freeze out, so a statement about the mechanism is conceivable. In~\cite{Abreu:2020jsl} we showed that the triangle's characteristic time is proportional to $1/\Gamma_1$, because it is the decay 
lag for $1\to 3+B$ that delays particle 3 from reaching particle 2, previously and collinearly emitted, and closing the triangle. In this process, with $\Gamma_1 \sim 7-15$ MeV $\sim 14-28$ fm$^{-1}$, the triangle is reasonably completed. Setting it to 20 fm$^{-1}$, the detailed computation yields $\tau_\triangleleft=10$ fm.

Thus, we conclude that this structure, as others that we have studied before,
can be distinguished in a heavy ion collision from a real new hadron, be it tetraquark, molecule or else.

\section*{Acknowledgement}

This work was partially supported by the following grants: 
spanish MINECO FPA2016-75654-C2-1-P, MICINN PID2019-108655GB-I00 and -106080GB-C21; 
EU’s 824093 (STRONG2020); and UCM’s 910309 group grants and IPARCOS, as well as
the Brazilian CNPq (contracts 308088/2017-4 and 400546/2016-7) and FAPESB (contract
INT0007/2016).

%%%%%%%%%%%%%%%%%%%%%%%%%%%%%%%%%%%%%%%%%%%%%%%%%%

\end{document}